\documentclass[a4paper]{jpconf}
\usepackage{graphicx}
\usepackage{amssymb}

\begin{document}
\title{New developments in the statistical approach of parton
distributions }

\author{Jacques Soffer}

\address{Centre de Physique Th\'eorique, UMR 6207 \footnote{UMR 6207 - Unit\'e Mixte 
de Recherche du CNRS et des Universit\'es Aix-Marseille I,
Aix-Marseille II et de l'Universit\'e du Sud Toulon-Var - Laboratoire 
affili\'e à la FRUMAM},\\
CNRS-Luminy, Case 907\\
F-13288 Marseille Cedex 9 - France }

\ead{soffer@cpt.univ-mrs.fr}

\begin{abstract}
We recall how parton distributions are
constructed in a statistical physical picture of the nucleon. 
The chiral properties of QCD lead to strong relations between quarks and 
antiquarks distributions and the importance of the Pauli exclusion principle
is also emphasized. A global next-to-leading order QCD analysis of unpolarized and polarized
deep-inelastic scattering data allows to determine a small number of
free parameters. Some predictions are compared to recent experimental results and we discuss
the prospects of this physical framework. 
\end{abstract}

\section{Introduction}
Deep-inelastic scattering (DIS) of leptons on hadrons has been extensively
studied, over the last twenty years or so, both theoretically 
and experimentally, to extract the polarized parton distributions (PPD) of the nucleon.
As it is well known, the unpolarized light quarks ($u,d$) distributions are fairly 
well determined. Moreover, the data exhibit a clear evidence for a
flavor-asymmetric light sea, {\it i.e.} $\bar d > \bar u$, which can be
understood in terms of the Pauli exclusion principle, based on the fact 
that the proton contains two $u$ quarks and only one $d$ quark. 
Larger uncertainties still persist for the gluon ($G$) and the heavy quarks 
($s,c$) distributions. From the more restricted amount of data on polarized 
structure functions, the corresponding polarized gluon and $s$ quark
distributions ($\Delta G, \Delta s$) are badly constrained and we just begin to
uncover a flavor asymmetry, for the corresponding polarized light sea, namely  
$\Delta \bar u \neq \Delta \bar d$. 
Whereas the signs of the polarized light quarks distributions are 
essentially well established, $\Delta u > 0$ and $\Delta d < 0$, 
this is not the case for $\Delta \bar u $ and $\Delta \bar d $.
In this talk, essentially based on Refs.\cite{BBS,BBS1,BBS2}, we will recall 
the construction of a complete set of polarized parton
(all flavor quarks, antiquarks and gluon) distributions and, in particular, 
we will try to clarify this last point on the polarized light sea.
Our main motivation is to use the statistical approach to
build up : $q_i$, $\Delta q_i$, $\bar q_i$, $\Delta \bar q_i$, $G$ and $\Delta
G$, by means of a {\it very small number} of free parameters. 
A flavor separation for the unpolarized and polarized light sea
is automatically achieved in a way dictated by our approach.
Our predictions are compared with some new 
unpolarized and polarized DIS measurements obtained at DESY, SLAC and
Jefferson Lab., as well as hadronic cross sections, from FNAL and
RHIC-BNL, and they turned out to be very satisfactory. 

\newpage

\section{Construction of the PPD in the statistical approach}

In the statistical approach the nucleon is viewed as a gas of massless partons 
(quarks, antiquarks, gluons) in equilibrium at a given temperature in a 
finite size volume. 
The light quarks $q=u,d$ of helicity $h=\pm$, at the input energy scale 
$Q_0^2=4\mbox{GeV}^2$, are given by the sum of two terms \cite{BBS},
a quasi Fermi-Dirac function and a helicity independent diffractive
contribution, common to all light quarks
\begin{equation}
xq^{h}(x,Q_0^2)= \frac{A X_{0q}^h x^b}{\exp[(x-X_{0q}^h)/{\bar x}]+1} +
\frac{{\tilde A} x^{\tilde b}}{\exp(x/{\bar x})+1}~.
\label{1}
\end{equation}
Here $X^{h}_{0q}$ is a constant, which plays the role of the 
{\it thermodynamical potential} of the quark $q^h$ and $\bar{x}$ is the 
{\it universal temperature}, which is the same for all partons.
Clearly one has $q_i= q_i^{+} + q_i^{-}$, for the unpolarized quark distributions
and $ \Delta q_i= q_i^{+} - q_i^{-}$ for the polarized ones
and similarly for antiquarks and gluons.
We recall that from the chiral structure of QCD, we have two important 
properties, allowing to relate quark and antiquark distributions and to
restrict the gluon distribution:

- The potential of a quark $q^{h}$ of helicity {\it h} is opposite to the
potential of the corresponding antiquark $\bar q^{-h}$ of helicity {\it -h}
\begin{equation}
X_{0q}^h=-X_{0\bar q}^{-h}~.
\label{2}
\end{equation}

- The potential of the gluon $G$ is zero
\begin{equation}
X_{0G}=0~.
\label{3}
\end{equation}
Therefore similarly to Eq.~(\ref{1}), we have for the light antiquarks
\begin{equation}
x\bar q^{h}(x,Q_0^2)= \frac{\bar A (X_{0q}^{-h})^{-1}
x^{2b}}{\exp[(x+X_{0q}^{-h})/{\bar x}]+1}
+\frac{{\tilde A}x^{\tilde b}}{\exp(x/{\bar x})+1}~.
\label{4}
\end{equation}
Here we take $2b$ for the power of $x$ and not $b$ as for quarks, 
an assumption which was discussed and partly justified in Ref.~\cite{BBS}.\\
Concerning the unpolarized gluon distribution, we use a quasi Bose-Einstein 
function, with zero potential
\begin{equation}
xG(x,Q_0^2)=\frac{A_Gx^{b_G}}{\exp(x/{\bar x})-1}~.
\label{5}
\end{equation}
This choice is consistent with the idea that hadrons, in the DIS regime, 
are black body cavities for the color fields. 
It is also reasonable to assume that for very small {\it x}, $xG(x,Q_0^2)$ 
has the same behavior as the diffractive contribution of the quark and
antiquark distributions in Eqs.~(1) and (4), so we will take 
$b_G=1+\tilde b$. We also need to specify the polarized gluon distribution and 
we take 
\begin{equation}
x \Delta G(x,Q_0^2) = 0~,
\label{6}
\end{equation}
assuming a zero polarized gluon distribution at the input energy scale $Q_0^2$
.\\
For the strange quarks and antiquarks {\it s} and  $\bar s $, given our poor
knowledge on their unpolarized and polarized distributions, we take 
\begin{equation}
xs(x,Q_0^2)=x \bar s(x,Q_0^2)= \frac{1}{4}[x \bar u (x,Q_0^2) + x \bar
d(x,Q_0^2)]~,
\label{7}
\end{equation}
and
\begin{equation}
x\Delta s(x,Q_0^2)=x \Delta \bar s(x,Q_0^2)= \frac{1}{3}[x \Delta \bar d
(x,Q_0^2) -x \Delta \bar u(x,Q_0^2)]~.
\label{8}
\end{equation}
This particular choice gives rise to a large negative $\Delta s(x,Q_0^2)$.
Both unpolarized and polarized distributions for the heavy quarks 
{\it c, b, t}, are set to zero at $Q_0^2=4\mbox{GeV}^2$ and for larger $Q^2$, they
are generated by the $Q^2$ evolution.

With the above assumptions, we note that the heavy quarks do not introduce any
free parameters, likewise the gluons, since the normalization constant 
$A_G$ in Eq.~(5) is determined from the momentum sum rule. 
Among the parameters introduced so far in Eqs.~(\ref{1}) and (4),
$A$ and $\bar{A}$ are fixed by the two conditions $u-\bar{u}=2$, $d-\bar{d}=1$.
Clearly these valence quark conditions are independent of 
$\tilde b$ and $\tilde A$, since the diffractive contribution cancels out.
Therefore the light quarks require only {\it eight} free parameters, the 
{\it four} potentials $X^+_{0u}$, $X^-_{0u}$, $X^+_{0d}$, $X^-_{0d}$, 
{\it one} universal temperature $\bar x$ and three more parameters $b$, $\tilde b$ and $\tilde A$.

From well established features of the $u$ and $d$ quark distributions
extracted from DIS data, we anticipate some simple relations 
between the potentials:

- $u(x)$ dominates over $d(x)$, therefore one expects 
$X^{+}_{0u} + X^{-}_{0u} > X^{+}_{0d} + X^{-}_{0d}$

- $\Delta u(x) > 0$, therefore $X_{0u}^+ > X_{0u}^-$

- $\Delta d(x) < 0$, therefore $X_{0d}^- > X_{0d}^+$~.

So $X_{0u}^+$ should be the largest thermodynamical potential and
$X_{0d}^+$ the smallest one.
In fact, as we will see below, we have the following ordering
\begin{equation}
X_{0u}^+ > X_{0d}^- \sim  X_{0u}^- > X_{0d}^+ ~.
\label{9}
\end{equation} 
This ordering leads immediately to some important consequences for quarks and 
antiquarks. 

First, the fact that $X_{0d}^- \sim  X_{0u}^-$, indicated in Eq.~(9), 
leads to
\begin{equation}
u^-(x,Q_0^2) \lesssim d^-(x,Q_0^2)~,
\label{10}
\end{equation}
which implies from our procedure to construct antiquark from quark 
distributions,
\begin{equation}
\bar u^+(x,Q_0^2) \gtrsim \bar d^+(x,Q_0^2)~.
\label{11}
\end{equation}
These two important approximate relations were already obtained in 
Ref.~\cite{BBS}, by observing in the data, the similarity in shape 
of the isovector structure functions
$2xg_1^{(p-n)}(x)$ and $F_2^{(p-n)}(x)$, at the initial energy scale, 
as illustrated in Fig.~1. For  
$2xg_1^{(p-n)}(x)$ the black circles are obtained by combining SLAC 
\cite{slace155a} and JLab \cite{JLab04} data.
The white circles, which extend down to the very low $x$ region, include the 
recent deuteron data from COMPASS \cite{COM05} combined with the 
proton data from SMC \cite{SMC}, at the measured $Q^2$ values of these two 
experiments. The agreement with the curve of the
statistical model is improved in this later case. The + helicity components
disappear in the difference $2xg_1^{(p-n)}(x)-F_2^{(p-n)}(x)$. Since this difference
is mainly non-zero for $0.01 < x < 0.3$, it is due to the contributions of
$\bar u^-$ and $\bar d^-$ (see Ref.~\cite{BBS}).

Second, the ordering in Eq.~(9) implies the following properties for 
antiquarks, namely:

i) $\bar d(x) > \bar u(x)$, the flavor symmetry breaking which also follows
from the Pauli exclusion principle, as recalled above. 
This was already confirmed by the violation of the Gottfried sum rule 
\cite{Gott,NMC}.

ii) $\Delta \bar u(x) > 0$ and  $\Delta \bar d(x) < 0$, which have not been
established yet, given the lack of precision of the polarized semi-inclusive 
DIS data, as we will see below. One expects an accurate 
determination of these distributions from the measurement of helicity 
asymmetries for weak boson production in polarized $pp$ collisions at RHIC-BNL 
\cite{BSSW}, which will allow this flavor separation.
By performing a next-to-leading order (NLO) QCD evolution of these parton 
distributions, we were able to obtain in Ref.~\cite{BBS}, a good description 
of a large set of very precise data on the following unpolarized 
and polarized DIS structure functions 
$F_2^{p, d, n}(x,Q^2), xF_3^{\nu N}(x,Q^2)$
and $g_1^{p, d, n}(x,Q^2)$, in a broad range of $x$ and $Q^2$,
in correspondance with the {\it eight} free parameters :
\begin{equation}
X_{0u}^{+} = 0.46128,~ X_{0u}^{-} = 0.29766,~X_{0d}^{-} = 0.30174,~X_{0d}^{+} 
= 0.22775 \, ,
\label{12}
\end{equation}
\begin{equation}
\bar x =0.09907,~ b = 0.40962,~\tilde b = -0.25347,~\tilde A =0.08318\, ,
\label{13}
\end{equation}
and three additional parameters, which are fixed by normalization conditions
\begin{equation}
A = 1.74938,~\bar A = 1.90801,~A_G = 14.27535 ~,
\label{14}
\end{equation}
as explained above. Note that the numerical values of the four potentials are 
in agreement with the ordering in Eq.~(9), as expected, and all
the free parameters in Eqs.~(12, \ref{13}) have been determined rather 
precisely, with an error of the order of one percent.

\begin{figure}[h]
\begin{minipage}{20pc}
\includegraphics[width=20pc]{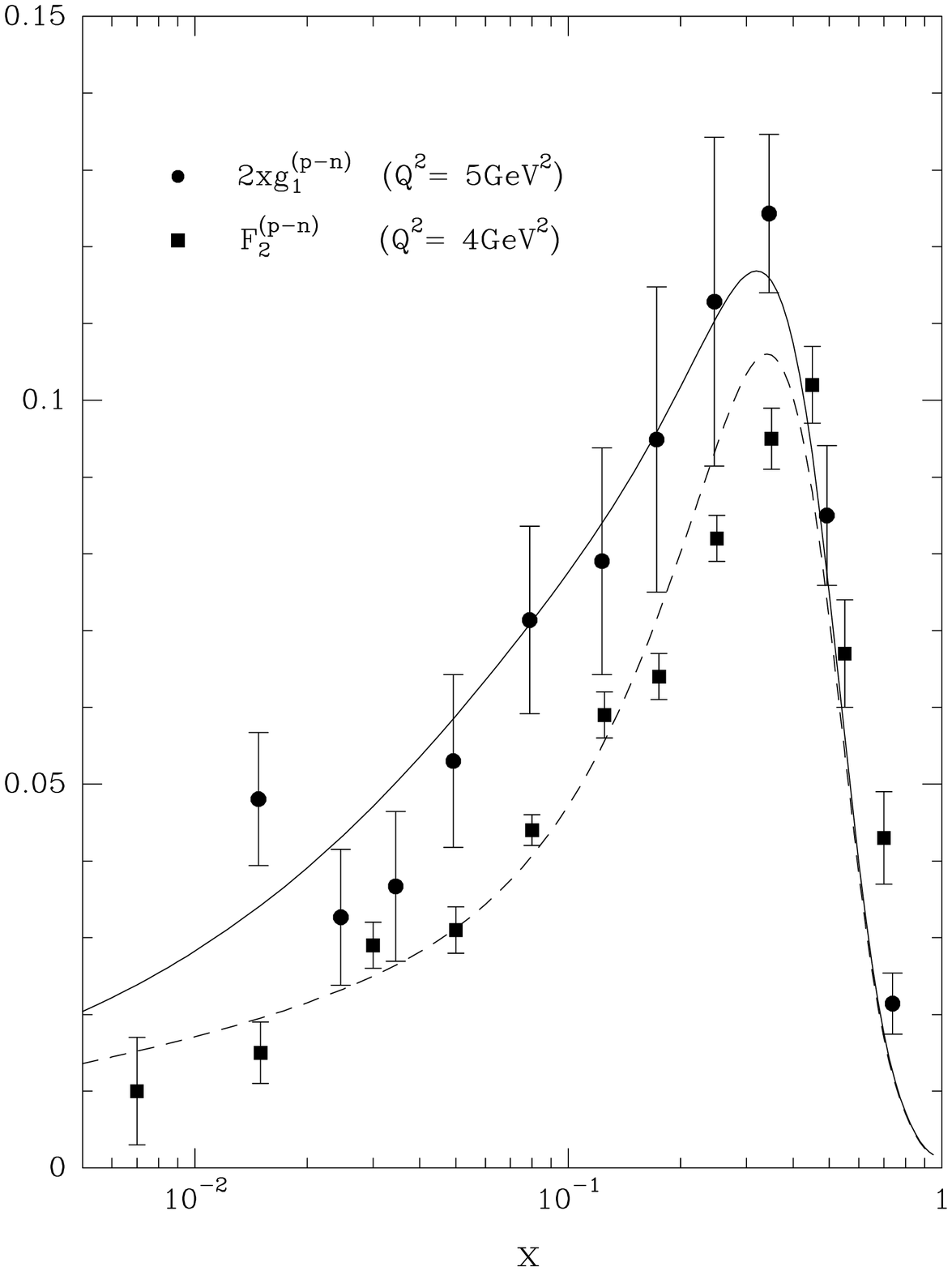}
\caption{\label{label}The isovector structure functions $2xg_1^{(p-n)}(x)$ (solid line from our 
statistical parton distributions) and $F_2^{(p-n)}(x)$
(dashed line from our statistical distributions).
Data are from  SLAC \cite{slace155a}, 
JLab \cite{JLab04}, COMPASS \cite{COM05}, SMC \cite{SMC} and NMC \cite{NMC} (Taken from Ref.~\cite{BBS2}).}
\end{minipage}\hspace{2pc}%
\begin{minipage}{18pc}
\includegraphics[width=18pc]{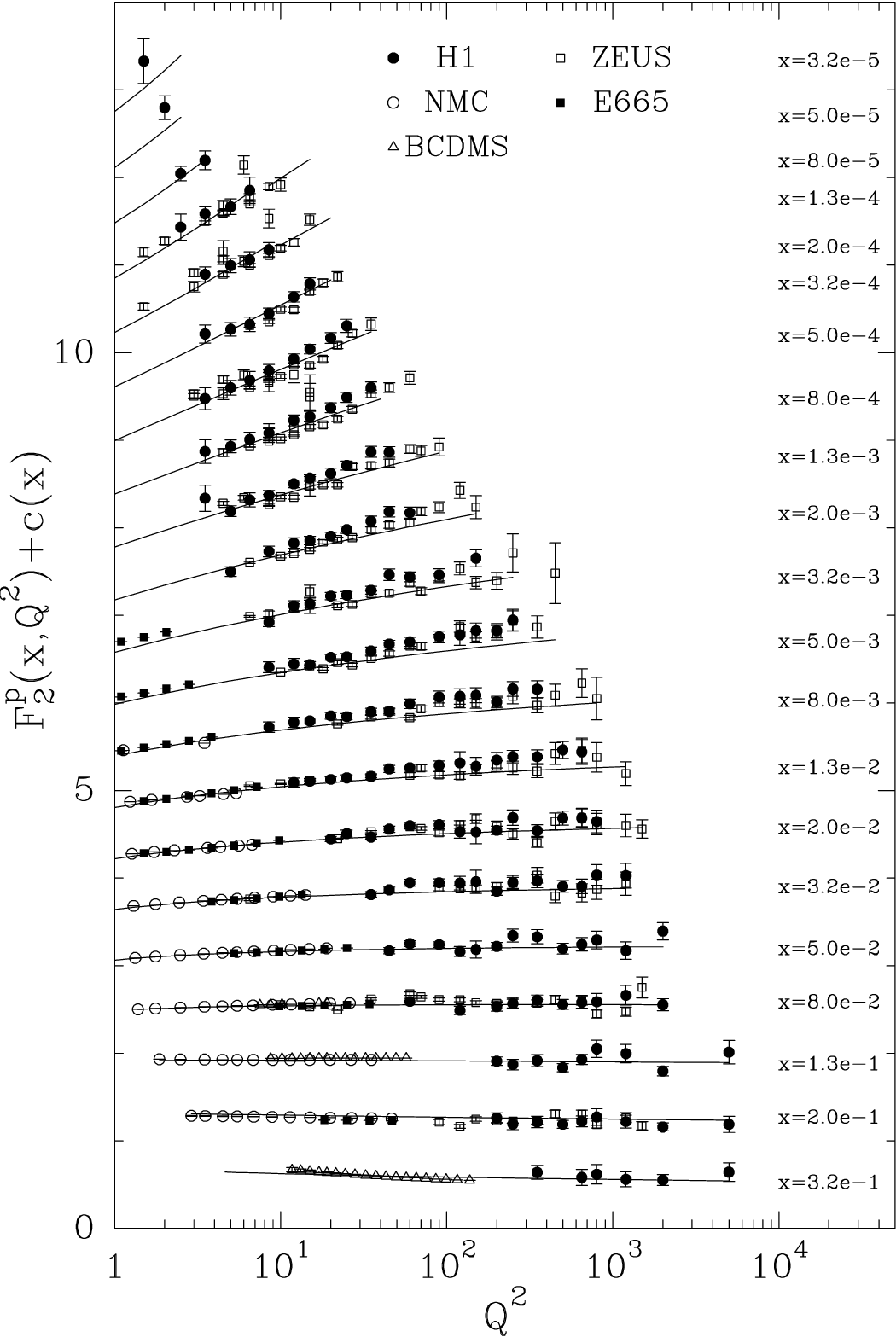}
\caption{\label{label}$F_2^p(x,Q^2)$ as function of $Q^2$ for fixed $x$,
$c(x) =0.6(i_x-0.4)$, $i_x = 1 \rightarrow x = 0.32$,
rebinned data H1, ZEUS, E665, NMC, BCDMS. (Presentation of data, courtesy of 
R. Voss) (Taken from Ref.~\cite{BBS})
.}
\end{minipage} 
\end{figure}

\section{Experimental tests for unpolarized and polarized DIS}
We first consider $\mu p$ and $ep$ DIS for which several experiments have
yielded a large number of data points on the structure function
$F_2^p(x,Q^2)$.
We have compared our calculations with fixed target
measurements NMC, BCDMS and E665, which cover a rather
limited kinematic region in $Q^2$ and also with the data at HERA from the H1
and ZEUS Collaborations.
These last data cover a very large $Q^2$ range, up to $Q^2=10^4\mbox{GeV}^2$
or so and probe the very low $x$ region which is dominated
by the rising behavior of the universal diffractive term.
We compare our results with the data on Fig.~2.
We also have a very good description of the neutron structure function
$F_2^n(x,Q^2)$ data, as well as for the  differential inclusive neutrino and antineutrino
cross sections from the high statistics $\nu N$ DIS, CCFR and NuTeV experiments as shown 
on Figs.~3 and 4 . As expected, for fixed $x$, the $y$ 
dependence is rather flat for neutrino and has the characteristic 
$(1-y)^2$ behavior for antineutrino.

\begin{center}
\begin{figure}[h]
\begin{minipage}{18pc}
\includegraphics[width=18pc]{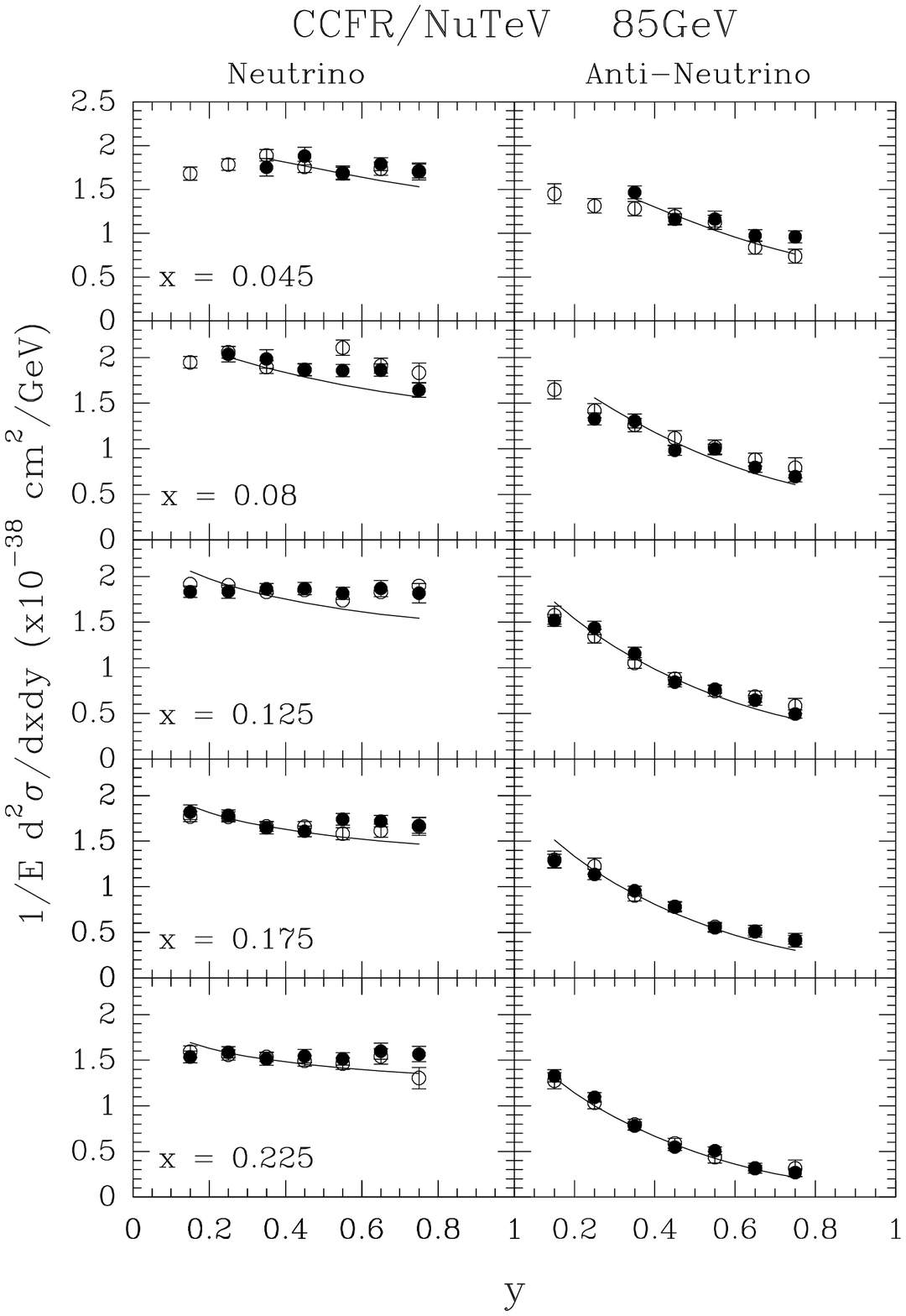}
\caption{\label{label}Differential cross section $\nu (\bar \nu )N$ for 
$E_{\nu} = 85 \mbox{GeV}$ and $0.045 \leq x \leq 0.225$, as
a function of $y$. Data are from CCFR \cite{UKYang} (white circles)
and NuTeV experiments \cite{nutevtalk,nutevdat} (black circles) (Taken from Ref.~\cite{BBS2}).}
\end{minipage}\hspace{2pc}%
\begin{minipage}{18pc}
\includegraphics[width=18pc]{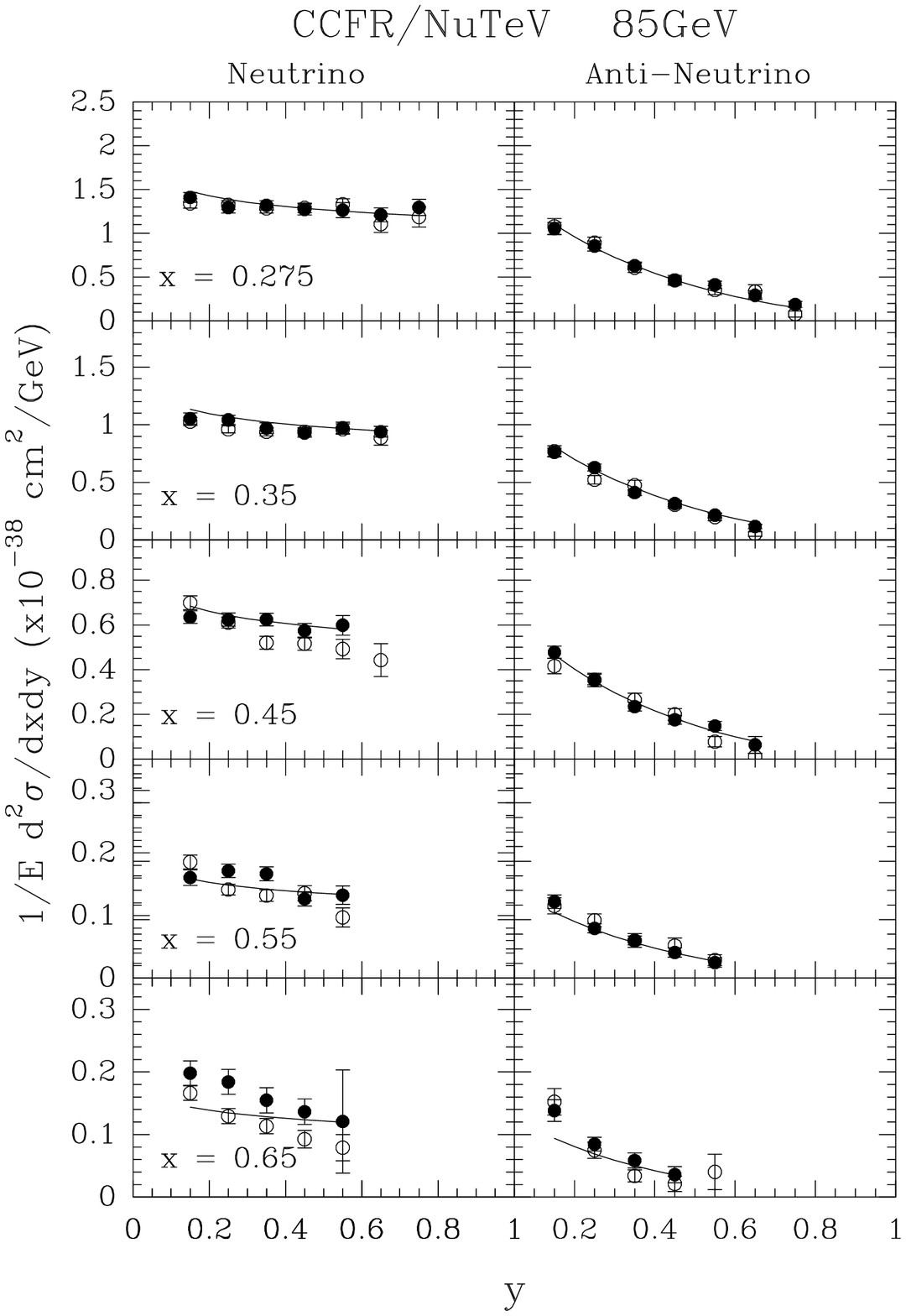}
\caption{\label{label}
Differential cross section $\nu (\bar \nu )N$ for 
$E_{\nu} = 85 \mbox{GeV}$ and $0.275 \leq x \leq 0.65$, as
a function of $y$. Data are from CCFR \cite{UKYang} (white circles)
and NuTeV experiments \cite{nutevtalk,nutevdat} (black circles) (Taken from Ref.~\cite{BBS2}).}
\label{fi:enu85}
\end{minipage} 
\end{figure}
\end{center}

Since our approach is based on the direct construction of the quark and
antiquark distributions of a given helicity  $q_i^{\pm}$ and $\bar q_i^{\pm}$, 
from the previous results we immediately obtained $\Delta q_i$ 
and $\Delta \bar q_i$ for each flavor, which enter in the definition of the
polarized structure functions
$g_1^{p,d,n}(x,Q^2)$.
In Fig.~5 we show a data compilation of 
$g_1^{p,d,n}(x,Q^2)$ from different current experiments on proton, 
deuterium and helium targets, evolved at a fixed value $Q^2 = 5\mbox{GeV}^2$. 
The $x$ dependence is in fair agreement with our results and we predict, in
the small $x$ region, a fast rising behavior for $g_1^p$ and a fast
decreasing behavior for $g_1^n$, due to the antiquark contributions. 
This cannot be tested so far, due to the lack of precise data. 
The very high $x$ region also remains poorly known and
the typical behaviour of the Fermi-Dirac functions, falling down
exponentially above the thermodynamical potential, which shows up in
Fig.~1, complies well with the fast change in the slope of
$g_1^p(x)$ at high $x$, as shown in Fig.~6.

\begin{center}
\begin{figure}[h]
\begin{minipage}{18pc}
\includegraphics[width=18pc]{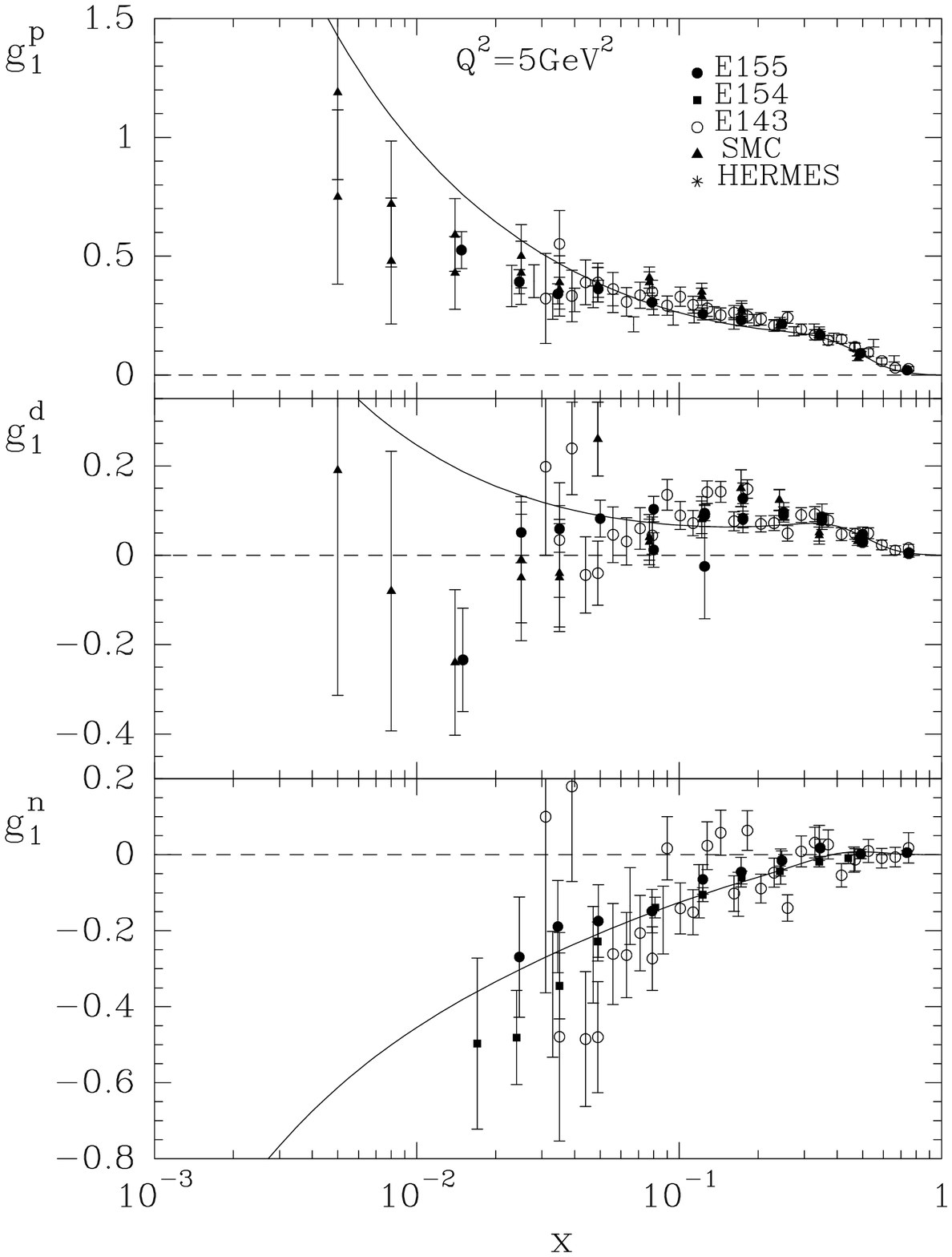}
\caption{\label{label}$g_1^{p,d,n}(x,Q^2)$ versus $x$ for different $Q^2$ values,
from E155, E154, E143, SMC, HERMES.
The curves correspond to our model predictions at $Q^2 = 5\mbox{GeV}^2$ (Taken from Ref.~\cite{BBS}).}
\end{minipage}\hspace{2pc}%
\begin{minipage}{18pc}
\includegraphics[width=18pc]{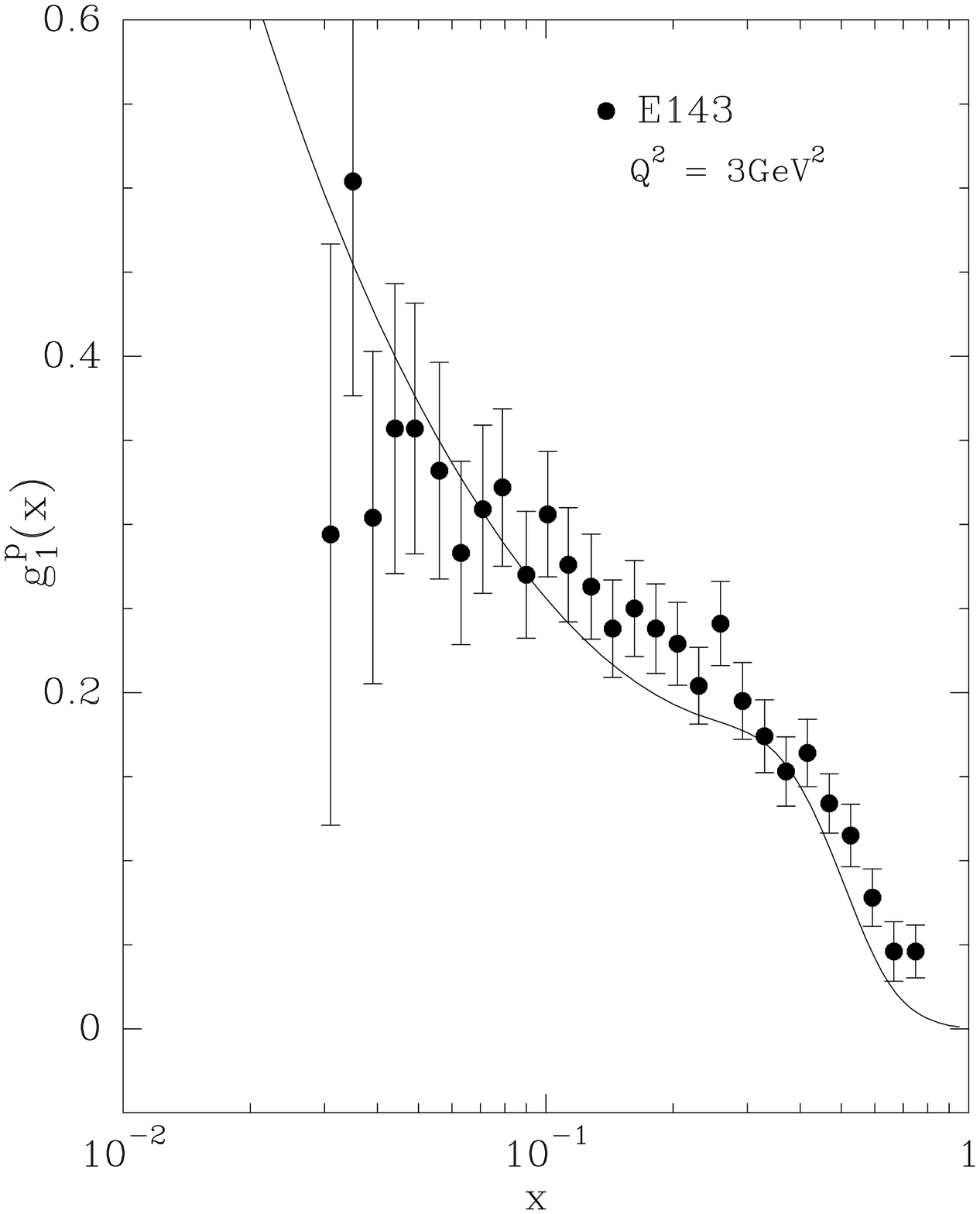}
\caption{\label{label}
$g_1^p(x,Q^2)$ as a function of $x$ at fixed $Q^2 = 3\mbox{GeV}^2$
from the statistical approach.
Experimental data from SLAC E143 \cite{slace143} (Taken from Ref.~\cite{BBS2}).}
\label{fi:g1}
\end{minipage} 
\end{figure}
\end{center}

Let us now turn to the helicity 
distributions, whose best determinations are shown in Figs.~7 and 8. 
The ratio $\Delta u(x)/u(x)$ 
is predicted to have a rather fast increase in the $x$ range 
$(X^-_{0u}-\bar{x},X^+_{0u}+\bar{x})$
and a smoother behaviour above, while $\Delta d(x)/d(x)$, which is negative,
has a fast decrease in the $x$ range $(X^+_{0d}-\bar{x},X^-_{0d}+\bar{x})$ 
and a smooth one above. This is exactly the trends displayed in 
Fig.~7 and our predictions are in perfect agreement
with the available accurate high $x$ data.
We note the behavior near $x=1$, another typical property of the statistical
approach, also at variance with predictions of the current literature. 
The fact that $\Delta u(x)$ is more concentrated in the higher $x$ region than
$\Delta d(x)$, accounts for the change of sign of $g^n_1(x)$, which becomes
positive for $x>0.5$, as first observed at Jefferson Lab \cite{JLab04a}.\\
Concerning the antiquark helicity distributions, it is clear that in
the very low $x$ region ($x<10^{-3}$) the ratio $\bar d(x)/\bar u(x)$ 
is $\sim 1$, because the diffractive contribution dominates. Since we predict
$\Delta \bar u(x) > 0$ and $\Delta \bar d(x) < 0$,
it is also interesting to remark that Eq.~(11) implies
\begin{equation}
\Delta \bar u(x) - \Delta \bar d(x) \simeq \bar d(x) - \bar u(x) > 0 ~,
\label{15}
\end{equation}
so the flavor asymmetry of the light antiquark distributions is almost the 
same for the corresponding helicity distributions. Similarly, 
Eq.~(10) implies
\begin{equation}
\Delta u(x) - \Delta d(x) \simeq  u(x) - d(x) > 0 ~.
\label{16}
\end{equation}
We also compare in Fig.~8 our predictions with an attempt from 
Hermes to isolate the different quark and antiquark helicity distributions. 
The poor quality of the data  does not allow to conclude on the signs of 
$\Delta \bar u(x)$ and $\Delta \bar d(x)$, which will have to wait for a
higher precision measurement of helicity asymmetries in $W^{\pm}$ production
at RHIC-BNL \cite{BS, BSa}.

\begin{center}
\begin{figure}[h]
\begin{minipage}{18pc}
\includegraphics[width=18pc]{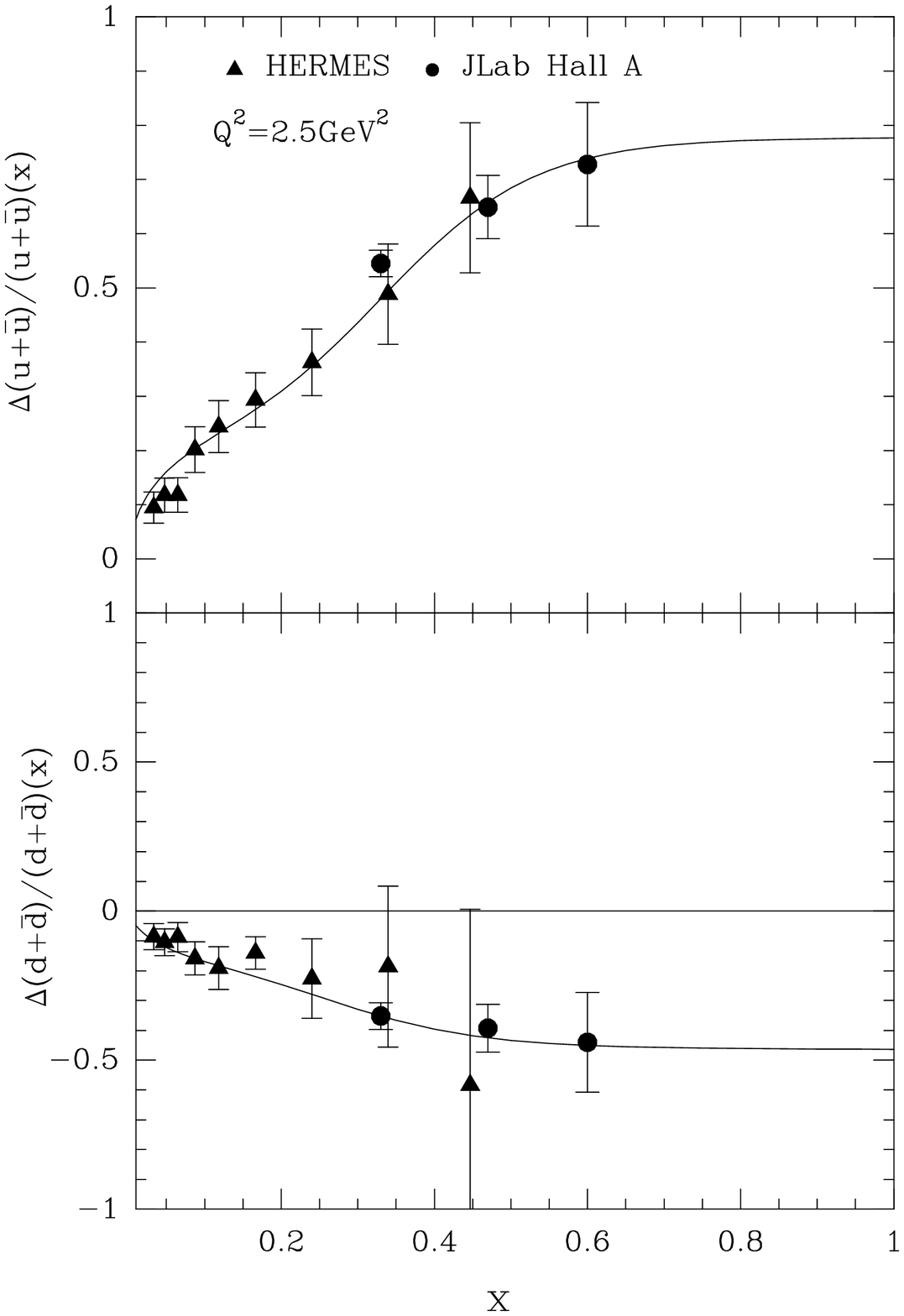}
\caption{\label{label}Ratios $(\Delta u + \Delta \bar u)/(u + \bar u)$ and 
$(\Delta d + \Delta \bar d)/(d + \bar d)$  as a function of $x$.
Data from Hermes for $Q^2 = 2.5\mbox{GeV}^2$ \cite{herm99} and
a JLab experiment \cite{JLab04a}. The curves are predictions from the 
statistical approach (Taken from Ref.~\cite{BBS2}).}
\end{minipage}\hspace{2pc}%
\begin{minipage}{18pc}
\includegraphics[width=18pc]{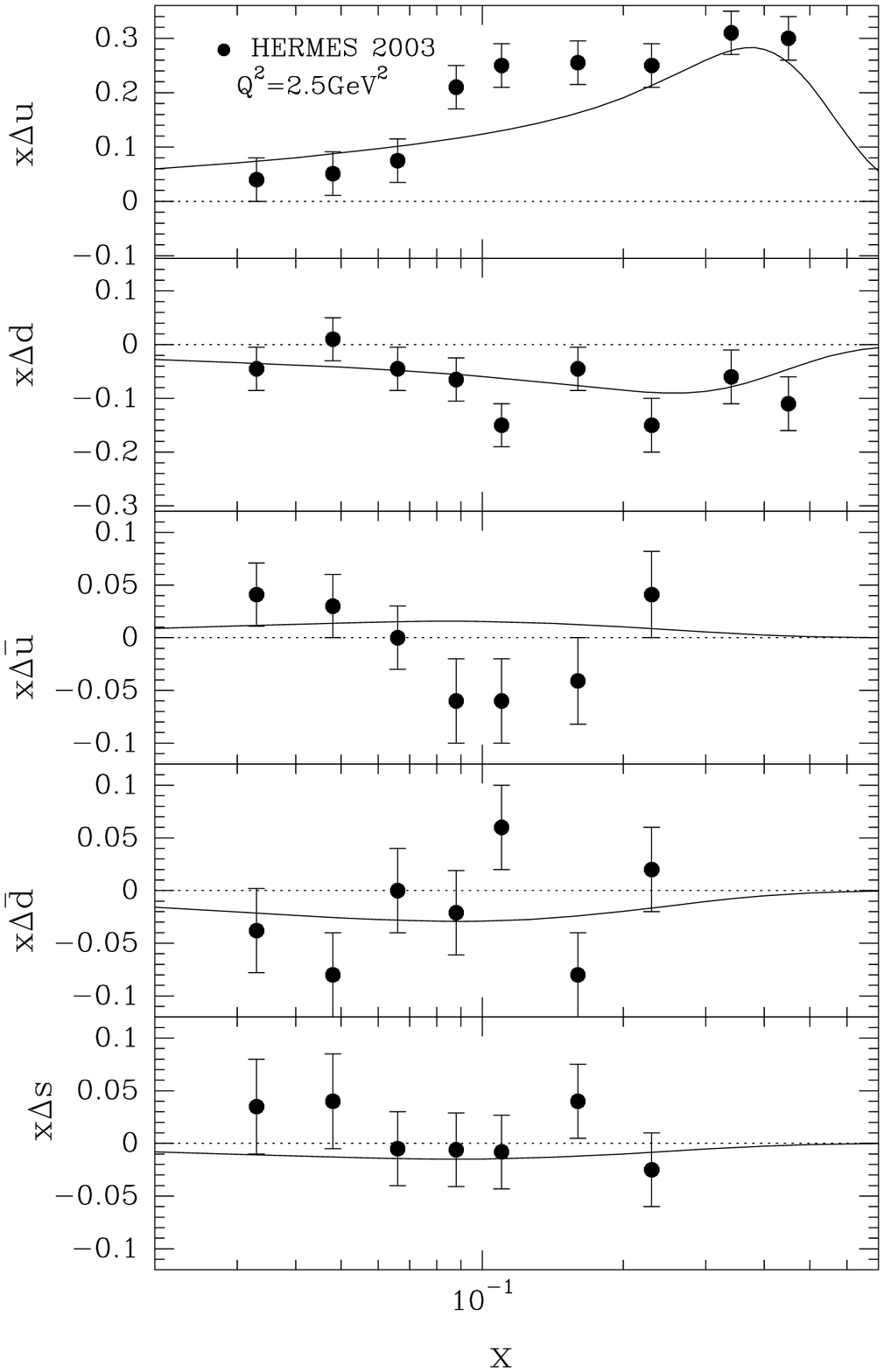}
\caption{\label{label}
Quarks and antiquarks polarized parton distributions as a function of $x$
for $Q^2 = 2.5\mbox{GeV}^2$. Data from Hermes \cite{herm04}. The curves are 
predictions from the statistical approach (Taken from Ref.~\cite{BBS2}).}
\label{fi:rap}
\end{minipage} 
\end{figure}
\end{center}

\section{Single jet and $\pi^0$ inclusive productions}
A precise determination of parton distributions allows us to use them
as input information to predict strong interaction processes, for 
additional tests of pertubative QCD and also for the search of
new physics. Here we shall test our statistical parton distributions for the  
description of two inclusive reactions, single jet and $\pi^0$ productions.
The cross section for the production of a single jet of rapidity $y$ and 
transverse momentum $p_T$, in a $\bar{p}p$ collision is given by  
\begin{eqnarray}
E\frac{d^3\sigma}{dp^3} 
&=& \sum_{ij}\frac{1}{1+\delta_{ij}}\frac{2}{\pi}
\int_{x_0}^{1} dx_a \frac{x_a x_b}{2x_a - x_Te^y} \times
\nonumber\\
&&\left[f_i(x_a, Q^2) f_j(x_b, Q^2)\frac{d\hat \sigma_{ij}}{d\hat{t}}(\hat s,
\hat t, \hat u) + (i\leftrightarrow j) \right]~,
\label{crossjet}
\end{eqnarray}
where $x_T = 2p_T / \sqrt{s}$, $x_0 = x_T e^y/(2 - x_T e^{-y})$,
$x_b = x_a x_T e^{-y}/(2x_a - x_T e^y)$ and $\sqrt{s}$
is the center of mass energy of the collision. In the above sum, $i,j$ stand 
for initial gluon-gluon, quark-gluon and quark-quark scatterings, 
$d\hat \sigma_{ij}/d\hat{t}$ are the corresponding partonic cross sections 
and $Q^2$ is the scaling variable.

In Fig.~9 our results are compared with the data from CDF and D0
experiments \cite{CDF2001,D02001}. Our prediction agrees very well with the 
data up to the highest $E_T$ (or $p_T$) value and this is remarkable given 
the fact that the experimental results are falling off over more than 
six orders of magnitude, leaving no room for new physics. For completeness,
we also show in Fig.~10 the D0 data, for several rapidity bins, using a
(Data-Theory)/Theory presentation.

\begin{center}
\begin{figure}[h]
\begin{minipage}{18pc}
\includegraphics[width=18pc]{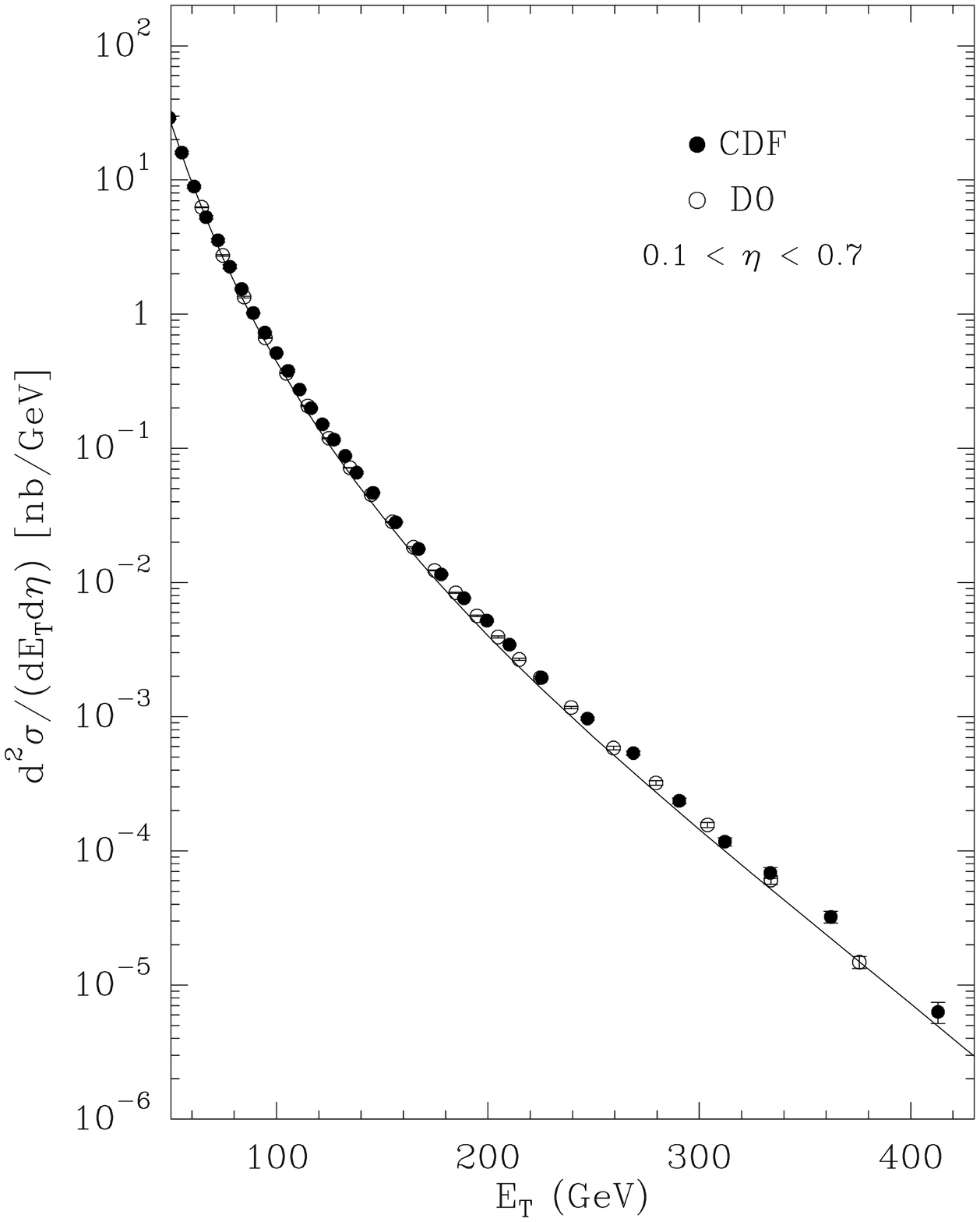}
\caption{\label{label}Cross section for single jet production in $\bar p p$ at $\sqrt{s} = 1.8
\mbox{TeV}$ as a function of $E_T$. Data are from CDF \cite{CDF2001}
and D0 \cite{D02001} experiments (Taken from Ref.~\cite{BBS2}).}
\end{minipage}\hspace{2pc}%
\begin{minipage}{18pc}
\includegraphics[width=18pc]{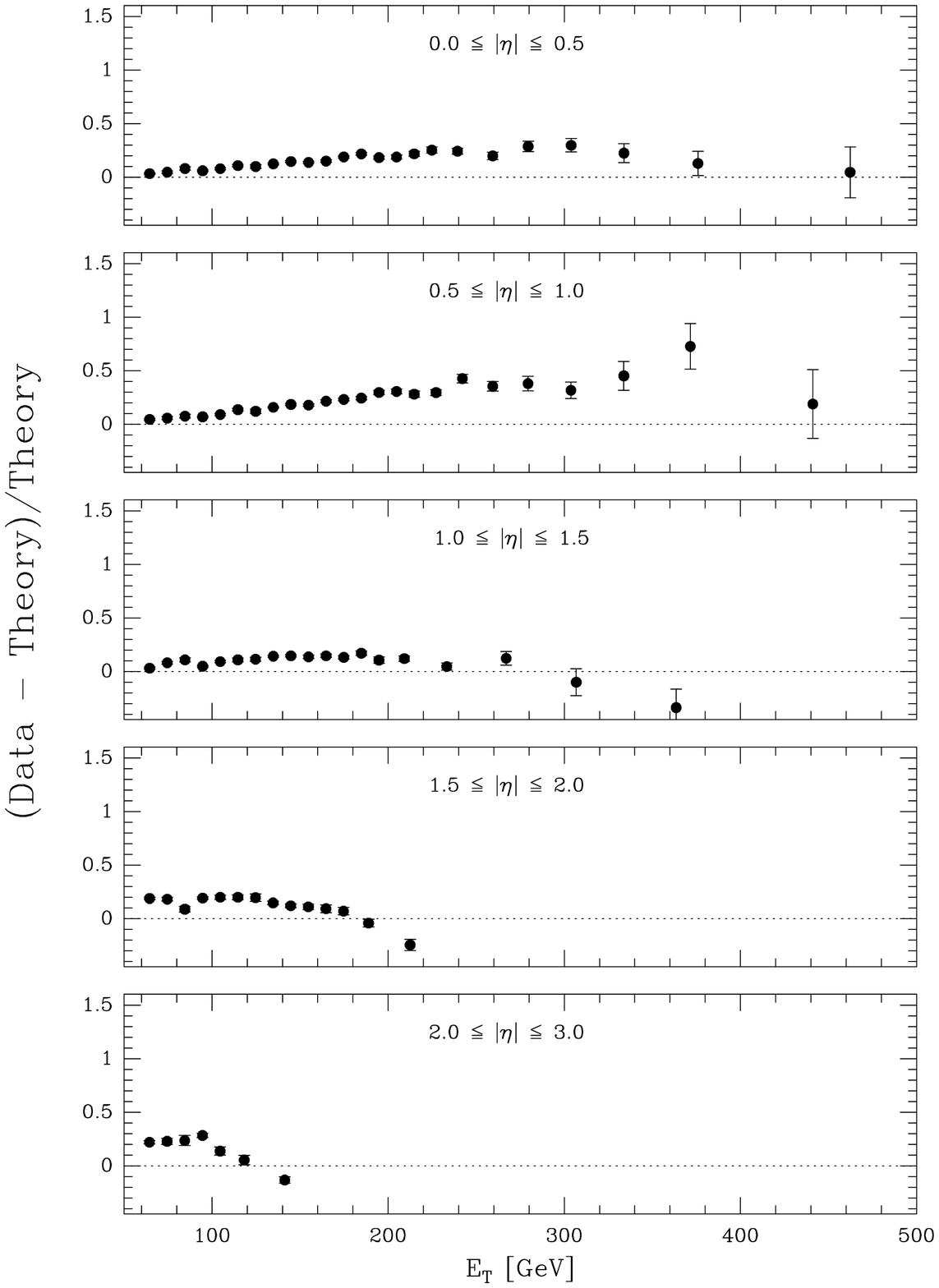}
\caption{\label{label}Comparison between the statistical model and the D0 \cite{D02001a}
single jet cross sections in $\bar p p$ at $\sqrt{s} = 1.8
\mbox{TeV}$, as a function of $E_T$ and rapidity $\eta$ (Taken from Ref.~\cite{BBS2}).}
\label{fi:jet}
\end{minipage} 
\end{figure}
\end{center}

Next we consider the cross section for the inclusive production of a $\pi^0$ 
of rapidity $y$ and transverse momentum $p_T$, in a $pp$ collision, which
has the following expression
\begin{eqnarray}
E_{\pi} d^3\sigma / dp_{\pi}^3
&=& \sum \limits_{abc} \int dx_a dx_b\, f_{a/p}(x_a,Q^2) \times  
\nonumber \\ 
&&f_{b/p}(x_b,Q^2)\frac{ D_{\pi^0/c}(z_c,Q^2)} 
{\pi z_c}d\hat {\sigma} /d \hat{t}(ab \to cX)~,
\label{Dsig}
\end{eqnarray}
where the sum is over all the contributing partonic channels $a b \to c X$ 
and $d \hat {\sigma}/ d \hat{t}$ is the associated partonic cross 
section. In these calculations $f_{a/p},f_{b/p}$ are our parton distributions
and $D_{\pi^0/c}$ is the pion fragmentation function.
Our calculations are done up to the NLO corrections, using the numerical code 
INCNLL of Ref. \cite{AVER} and for two different choices of fragmentation 
functions namely, BKK of Ref. \cite{binew95} and KKP of Ref. \cite{kniel00}, 
and we have checked that they give similar numerical results. We have compared
our predictions to two different data sets at $\sqrt{s} = 200\mbox{GeV}$
from PHENIX and STAR at RHIC-BNL. The results are shown in 
Figs.~11 and 12 and the agreement is good, both 
in central rapidity region (PHENIX) and in the forward region (STAR). This energy is
high enough to expect NLO QCD calculations to be valid in a large rapidity region, which
is not the case for lower energies \cite{BS0}.
\begin{center}
\begin{figure}[h]
\begin{minipage}{17pc}
\includegraphics[width=17pc]{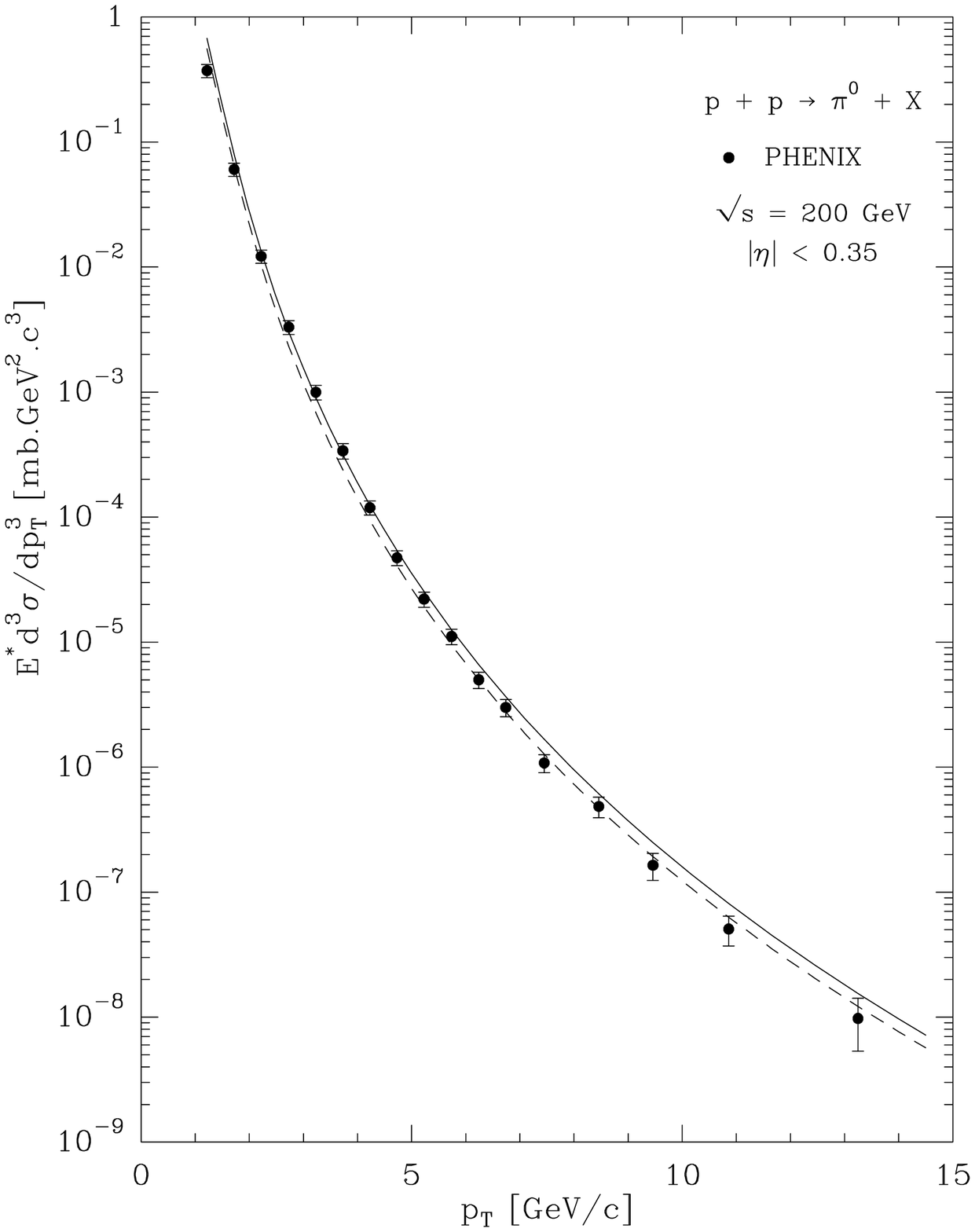}
\caption{\label{Label}Inclusive $\pi^0$ production in $p p$ reaction at $\sqrt{s} = 200
\mbox{GeV}$ as a function of $p_T$, scale $\mu = p_T$. 
Data from PHENIX \cite{phenix03}. Solid curve fragmentation functions from KKP \cite{kniel00},
dashed curve from BKP \cite{binew95} (Taken from Ref.~\cite{BBS2}).}
\end{minipage}\hspace{2pc}%
\begin{minipage}{17pc}
\includegraphics[width=17pc]{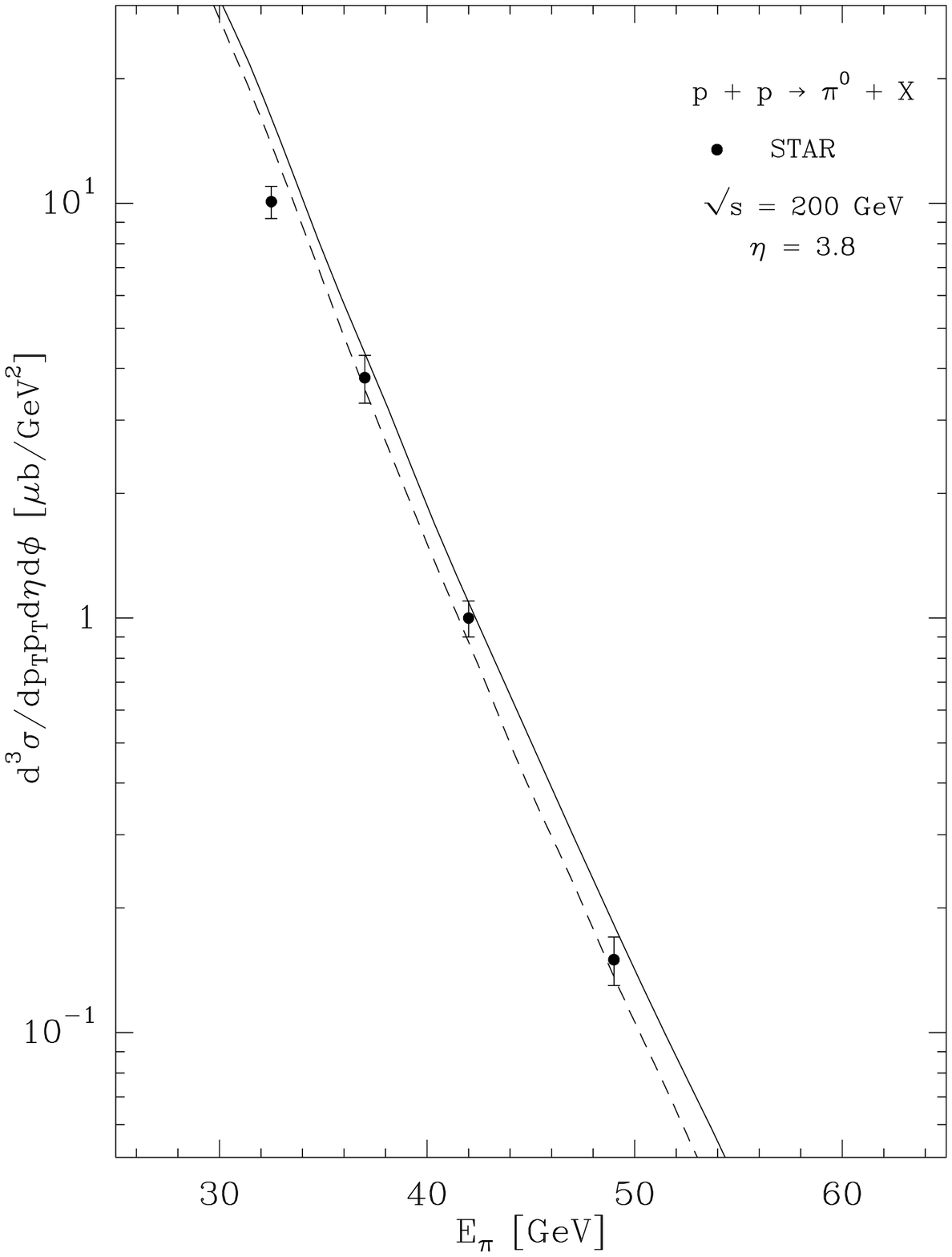}
\caption{\label{Label}Inclusive $\pi^0$ production in $p p$ reaction at $\sqrt{s} = 200
\mbox{GeV}$ as a function of $E_{\pi}$. Data from STAR \cite{star03}. 
Solid curve fragmentation functions from KKP \cite{kniel00},
dashed curve from BKP \cite{binew95} (Taken from Ref.~\cite{BBS2}).}
\label{fi:jet}
\end{minipage} 
\end{figure}
\end{center}
\newpage
\section{Final remarks}
We have shown that this simple approach of the statistical parton distributions,
which involves only eight free parameters, provides
a good description of recent data on unpolarized and polarized DIS 
and on some hadronic processes. Our distributions
have special features, which remain to be tested by new precise 
experimental data in the future, as well as its natural extension to
the transverse degree of freedom \cite{BBS3}. 

\ack
The author would like to thank the organizers of CORFU2005, a very successful meeting
in a most pleasant atmosphere. He is also very
grateful to Prof. George Zoupanos for the invitation
and for providing some financial support.

\end{document}